\documentclass[final,5p]{elsarticle}

\usepackage[utf8]{inputenc}
\usepackage{graphicx}

\usepackage[mathlines]{lineno}
\usepackage[version=4]{mhchem}

\usepackage{booktabs}
\usepackage{tabulary}
\usepackage{adjustbox}
\usepackage[hyphens]{url}

\usepackage{lineno}
\usepackage{setspace}
\usepackage{amssymb}

\usepackage{enumitem}
\usepackage{float}

\usepackage{wasysym}
\usepackage{siunitx}
\usepackage{xspace}
\usepackage{multirow}
\usepackage[normalem]{ulem}
\usepackage[colorlinks=true]{hyperref}
\usepackage[usenames, dvipsnames, table]{xcolor} 
\usepackage{placeins}

\usepackage{bm}

\biboptions{sort&compress}

\title{Replacement of a Photomultiplier Tube with Silicon Photomultipliers for use in Safeguards Applications}

\author[llnl]{Tyana Stiegler\corref{cor1}}
\ead{stiegler1@llnl.gov}
\author[llnl]{Kareem Kazkaz}
\ead{kazkaz1@llnl.gov}
\author[llnl]{Erik Swanberg}
\ead{swanberg3@llnl.gov}
\author[llnl]{Vladimir Mozin}
\ead{mozin1@llnl.gov}
\address[llnl]{Lawrence Livermore National Laboratory, Livermore, CA, USA}
\date{\today}
\cortext[cor1]{Corresponding author \textit{stiegler1@llnl.gov}}

\begin{document}
\begin{abstract}
 We compared the performance of a SiPM array and a PMT in a laboratory setting using a single 5.08$\times$5.08-cm cylindrical sodium iodide scintillating crystal. Photomultiplier tubes (PMTs) are the most commonly used device to monitor scintillating materials for radiation detection purposes. The systems are sometimes limited by disadvantages in the PMTs that may degrade their performance, including temperature dependence and variation with magnetic field. Instrumentation engineering must also contend with a potentially large volume relative to the active scintillator volume, fragility, and high voltage requirements. One possible alternative is an array of silicon photomultipliers (SiPMs).  Measurements were made with a 5.04$\times$5.04-cm sensL J-series SiPM array and a 7.62~cm Hamamatsu PMT. We demonstrated how the SiPM bias can be sufficiently altered to remove the effects of temperature variation encountered in environments where nuclear safeguards work is often performed. Finally, we evaluated a method of determining enrichment levels of \ce{^235U} at various levels and shielding configurations, using both the PMT-mounted and SiPM-mounted scintillator.
\end{abstract}


\begin{keyword}

Silicon Photomultipliers \sep NaI(Tl) \sep Gamma-ray Spectrometry \sep Uranium \sep Energy Resolution \sep Nuclear Monitoring
\end{keyword}

\maketitle


\section{\label{sec:intro}Introduction}
    
Inorganic scintillation detectors are widely used in gamma ray spectroscopy, as they are available at low cost and large size, have relatively high gamma stopping power, and have sufficient energy resolution for a variety of use scenarios. A very common spectroscopy system is a thallium-doped sodium iodide (NaI) crystal instrumented with a photomultiplier tube (PMT). Hand-held versions of these systems are important tools for nuclear safeguards, first responders, and in the prevention of illicit trafficking of nuclear materials~\cite{gammaRad5,inspec1000,identR400}. Over decades of use, engineers and scientists have identified a number of disadvantages of PMTs. The level of concern of each depends on the application and environment.

Typical disadvantages cited include bulkiness, fragility, susceptibility to magnetic fields, and high voltage requirements (typically $\gtrsim$1000~V)~\cite{RFWireless,Faham2009,Laserfocus,Barbarino2014}.
Emerging technologies could mitigate these disadvantages while maintaining parity with the performance and cost of a PMT. One of these alternatives is the silicon photomultiplier (SiPM), which has several aspects that could make them preferable to a PMT. They are compact, no not require a vacuum volume, are insensitive to magnetic fields, run at low bias voltages (30-100~V), are physically robust, and are comparable in price to a PMT. SiPM response curves are more dependent on temperature, though, an aspect that we address later in this work.

The goal of this experiment was to asses the viability of replacing a 7.62~cm Hamamatsu  PMT with a 5.04$\times$5.04-cm sensL J-series SiPM array in a typical hand held spectrometer. These photodetectors' active areas were larger than the dimension of the scintillator, ensuring maximal light collection. Comparisons were carried out by measuring the FWHM energy resolution at several energies, and exploring temperature dependence and possible stabilization methods. We then compared the performance of each photodetector using several \ce{^235U} enrichment standards by measuring the energy resolution of the \ce{^235U}-186~keV and \ce{^238U}-1001~keV gamma peaks, as well as the enrichment predictive capability.

This study did not include investigation of magnetic field effects but this has been reported on in other experiments \cite{sipmMag1,sipmMag2,sipmMag3}.

The following sections detail the experimental setup and results of our comparison. Section~\ref{sec:expSetup} describes the physical details, the calibration, and resulting energy resolution measurements. The effects of varying temperature and how to compensate is detailed in Section~\ref{sec:temp}. Section~\ref{sec:Usens} presents results of the \ce{^235U} enrichment standards campaign.


\section{\label{sec:expSetup}Experimental Details and Energy Calibration}

Details of the hardware used in these evaluations are given in Table~\ref{tab:PrimaryComponents}. Each photodetector was mounted in turn to the same NaI scintillator to avoid systematic effects from using different crystals. Each photodetector was chosen to ensure full coverage of the NaI, for good light collection. Optical grease was used to mount the photodetectors, again to maximize detection efficiency. We selected the sensL ArrayX-BOB6-64S SiPM readout board because it sums over all pixels, allowing for single-channel readout of the device. This allowed the back-end electronics and analysis nearly identical to that of the PMT, with a signal polarity flip and a slight gain adjustment on the amplifier being the only alterations.

\begin{figure}[tbp]
        \centering
        \includegraphics[width=0.9\columnwidth]{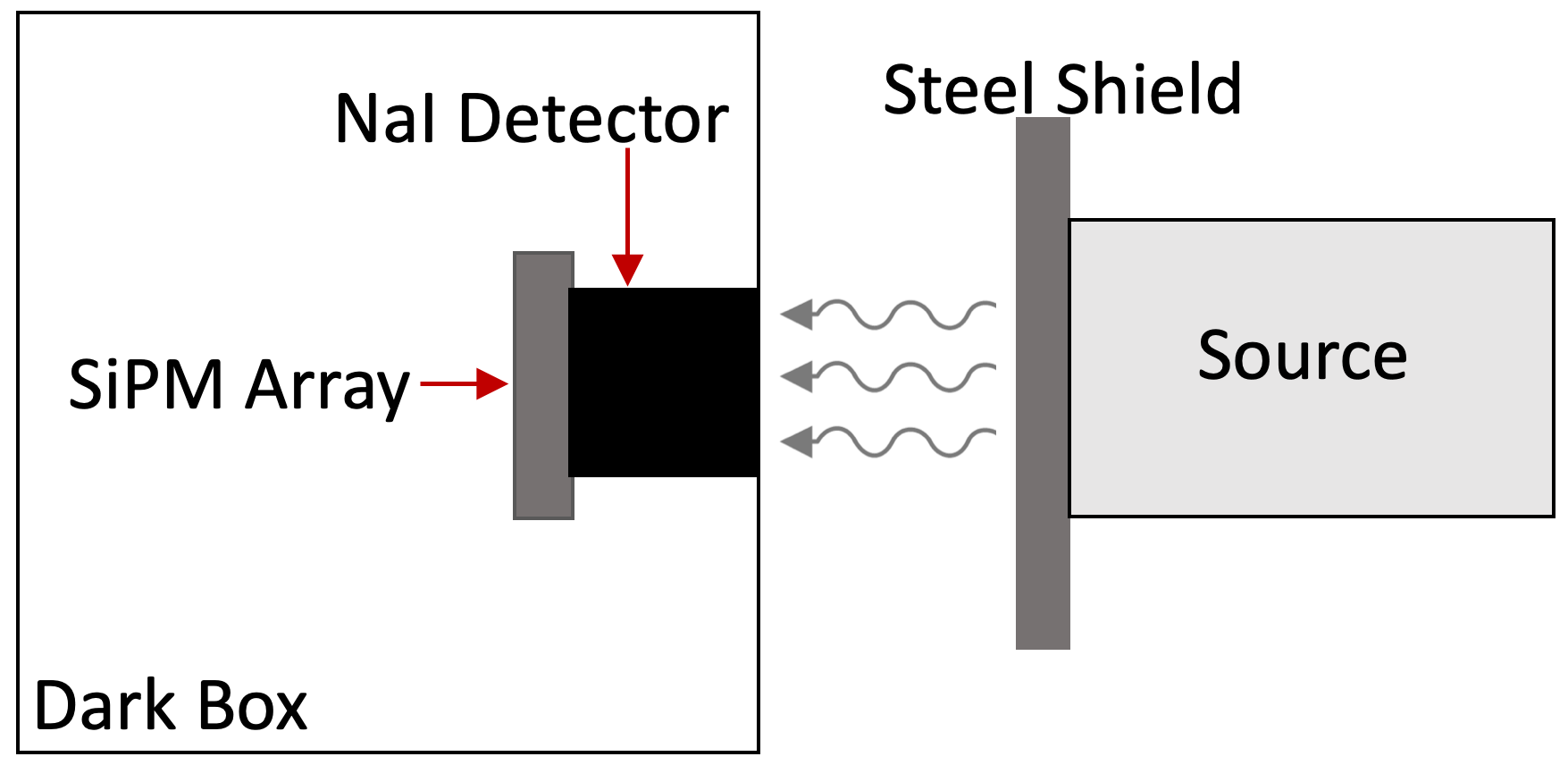}
        \caption{Schematic of one configuration of the experiment: the SiPM-mounted NaI detector inside the dark box. The PMT-mounted data was acquired by replacing the SiPM array, but keeping the crystal in the same position inside the box. The steel shield was used as part of the uranium data campaign, and was removed for all background and calibration datasets.}
        \label{fig:schematic}
\end{figure}

A schematic of the experimental setup is shown in Fig.~\ref{fig:schematic}. When the SiPM and PMT were exchanged, we took care to position the crystal, dark box, and sources in consistent locations to minimize effects of solid angle coverage, backscatter, or intervening material. The steel shield was used in the uranium campaign and was not present for the energy calibrations.

\begin{figure}[tbp]
        \centering
        \includegraphics[width=0.9\columnwidth]{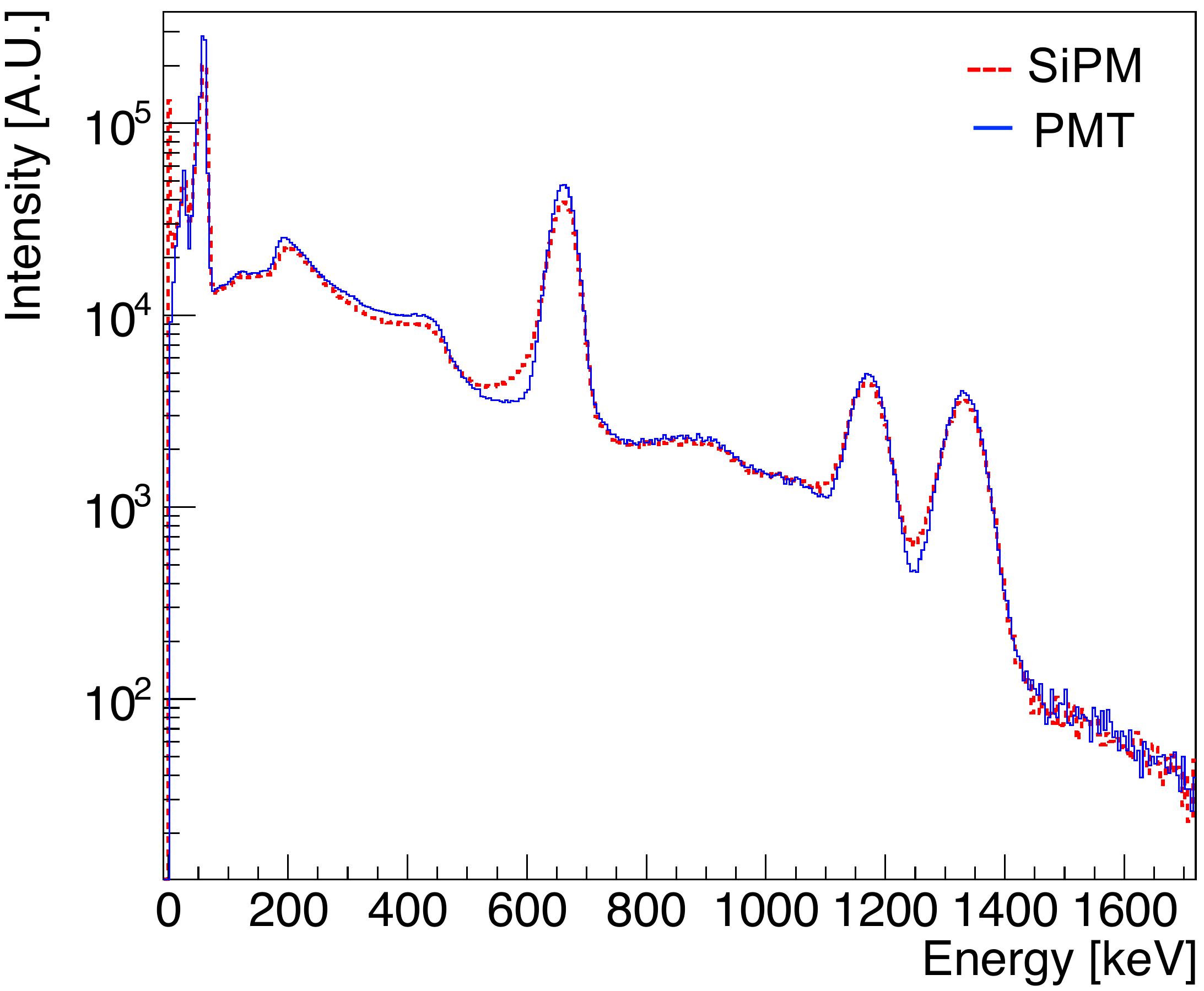}
        \caption{Full spectrum from calibration sources \ce{^241Am}, \ce{^137Cs}, and \ce{^60Co}. Solid Blue: PMT, Dashed Red: SiPM}
        \label{fig:FullSpec}
\end{figure}

For each photodetector, we acquired background spectra as well as data from three calibration sources: \ce{^241Am} (59.5~keV), \ce{^137Cs} (662~keV). and \ce{^60Co} (1173~ and 1332~keV). The background and calibration sets were taken multiple times during the uranium measurements to ensure stability of the detector response. A typical calibration spectrum before background subtraction is shown in Fig.~\ref{fig:FullSpec}. The background spectrum was subtracted from all datasets before analysis. The calibration sources were chosen to provide gamma rays that bracket the energy range of gammas of interest from \ce{^235U} and \ce{^238U}. The fit function to characterize the resolution of the detectors is a Gaussian curve over an inverted Heaviside function.  

\begin{table*}[htb]
    \centering
    \begin{tabular}{llll}
    \hline
    \hline
    \textbf{Component} & \textbf{Manufacturer} & \textbf{Model} & \textbf{Description}\\
    \hline
                     &              &          & 5.08$\times$5.08-cm cylindrical NaI\\        
                     &              &          & crystal packaged in air-tight \\
        NaI detector & Saint Gobain & SA-12428 & aluminum housing with a glass \\
                     &              &          & window and reflective internal\\  
                     &              &          & wrapping\\
    \hline
                     &              &                 & 5.04$\times$5.04-cm, 8$\times$8 pixel array,\\
                     &              &                 & with summed breakout electronics\\
        SiPM array   & sensL        & J-Series 60035  & board. Each pixel is 6~mm on a \\                       &              &                 & side. The single-channel readout\\
                     &              &                 & board was an ArrayX-BOB6-64S. \\            
    \hline
                     &              &               & 7.62~cm bialkali photocathode  \\
         PMT         & Hamamatsu    & R6233-100 SEL & and borosilicate glass \\
                     &              &               & window \\
    \hline
    \hline
    \end{tabular}
        \caption{Primary components used in the laboratory comparison. The output from both the PMT and SiPM were connected to a multichannel analyzer to record the spectra. The breakout board for the SiPM allowed the 64 pixels to be read out as a single summed channel.}
        \label{tab:PrimaryComponents}
\end{table*}

 
The resolution of the NaI mounted to each photodetector is shown in Table~\ref{tab:res}. Resolution is in part a function of the number of detected photons. The resolution at low energies of the SiPM array is degraded relative to that of the PMT because SiPMs have high dark count rates while PMTs are very low noise devices. Modern SiPMs have higher light collection efficiency which can produce better resolution than PMTs at high energies. Other effects could be electronic noise or the non-linearity in the SiPM response. Further investigation into the resolution in this specific configuration is reserved for a future study.

\begin{table*}[tbh!]
        \centering
        \begin{tabular}{lcccc}
        \hline
        \hline
        \textbf{Detector} & \textbf{Size} & \textbf{Active Area}& \textbf{Quantum Eff. or}\\
        \textbf{} & \textbf{$[cm]$} & \textbf{$[cm^{2}]$} & \textbf{Photon Det. Eff.}\\
        \hline
            PMT & 7.62 round & 20.3 & 30\%\\
            SiPM & 5.04 square & 14.4 & 50\% \\
        \hline
        \hline
        \end{tabular}
        \caption{Specifications of the PMT and the SiPM array. Efficiency for the PMT and SiPM is in quantum efficiency and single photon detection efficiency respectively. Note that it is not the full area of the PMT and SiPMs that are used, but the overlap of the photodetectors with the 5.08~cm NaI crystal. 
        }
        \label{tab:specs}
\end{table*}

\begin{table*}[tbh!]
        \centering
        \begin{tabular}{ccc}
        \hline
        \hline
        \textbf{Peak Energy} & \textbf{PMT Resolution} & \textbf{SiPM Resolution}\\
        \textbf{$[keV]$} & \textbf{$[\%]$} & \textbf{$[\%]$}\\
        
        \hline
            59 & $10.5\pm0.11$ & $14.13\pm0.14$\\
            662 & $6.72\pm0.03$ & $7.08\pm0.03$\\
            1332 &$5.00\pm0.03$ & $5.26\pm0.04$ \\
        \hline
        \hline
        \end{tabular}
        \caption{The full-width at half-maximum (FWHM) energy resolution of the NaI crystal with the PMT and SiPM array for the three calibration sources \ce{^241Am}, \ce{^137Cs}, and \ce{^60Co}.}
        \label{tab:res}
\end{table*}

\section{\label{sec:temp}Varying Bias to Compensate for Temperature Change}

The light output of sodium iodide crystals is known to exhibit a temperature dependence~\cite{Elmore1949,Reeder2004}, which the manufacturer characterizes as -0.3\%/$^\circ$C\cite{SaintGobain2016}. Given a temperature change from 24$^\circ$C to 0$^\circ$C, a preset detector calibration would have a deviation of 7\%, which is comparable to the FWHM resolution of the detector. This offset is sufficiently strong to give spurious results if the analysis does not take the temperature variation into account.

SiPMs themselves also display a temperature dependence independent of the scintillator. Given the mass and heat capacity differences between the NaI crystal and SiPM array, the components are not guaranteed to be in thermal equilibrium in the event of short-time-scale temperature cycling of the sort that regularly occurs in the field (e.g., warm storage location to cold car trunk to hot power plant chamber). This time-dependent temperature variation can lead to a complicated hysteresis  that hampers attempts to predict the response of the system as a whole. The bias applied to a SiPM, however, can be used to change the amplitude of its response. It is therefore possible in principle to compensate for temperature deviations once the system has come to thermal equilibrium.

\begin{figure}[tbp]
        \centering
        \includegraphics[width=0.9\columnwidth]{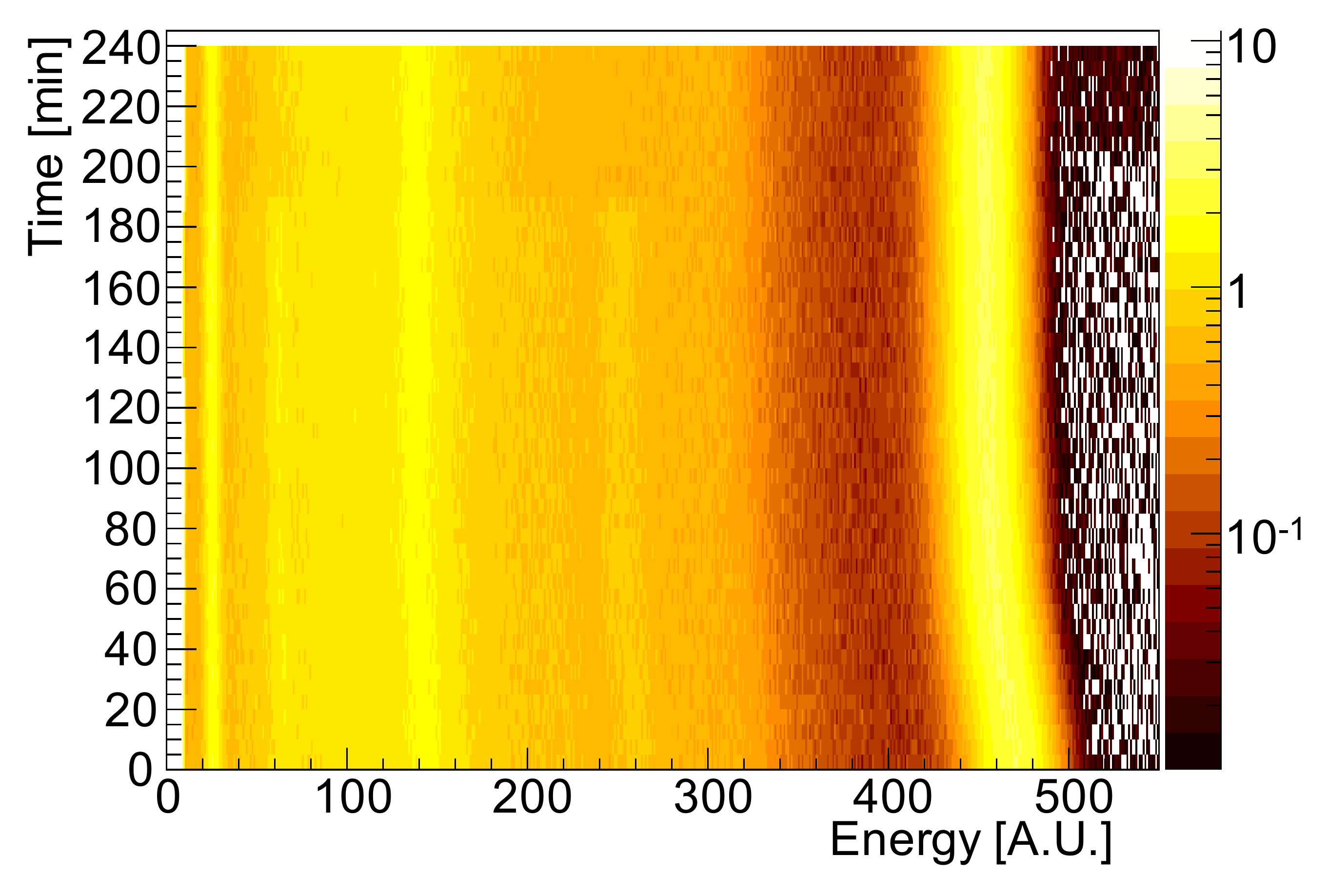}
        \caption{Uncalibrated $^{137}$Cs data with temperature variation acquired with the SiPM-mounted NaI. The system started at 21$^\circ$C, and stabilized at 27$^\circ$C. The system stabilized from this 6$^\circ$C temperature change after two hours. The Z axis shows intensity with arbitrary units. The signals at 60 and 260 on the energy axis that disappear at $\sim$190~minutes are from a $^{133}$Ba source that was close enough for the detector to observe before personnel put the source back in the source locker.}
        \label{fig:Waterfall}
\end{figure}

As part of our laboratory comparison, we explored the temperature and bias dependence of the SiPM array. If the SiPM demonstrates a dynamic range in the bias response sufficiently large to compensate for extreme, but realistic, temperature variations that are encountered in the field, it strengthens its viability as a replacement for PMTs in safeguards applications. The exploration begins with a characterization of the thermal equilibration time of the system. We put the SiPM-mounted NaI detector in an insulated environmental chamber at 21$^\circ$C, and began a series of calibration datasets with the $^{137}$Cs source. We turned on a hot plate inside the chamber, which gradually increased the temperature to 27$^\circ$C. Each $^{137}$Cs dataset was five minutes. The equilibration was measured over the course of four hours to determine the time to reach thermal equilibrium. Fig.~\ref{fig:Waterfall} shows the results, where the system stabilized after about two hours.

\begin{figure}[tbp]
        \centering
        \includegraphics[width=0.9\columnwidth]{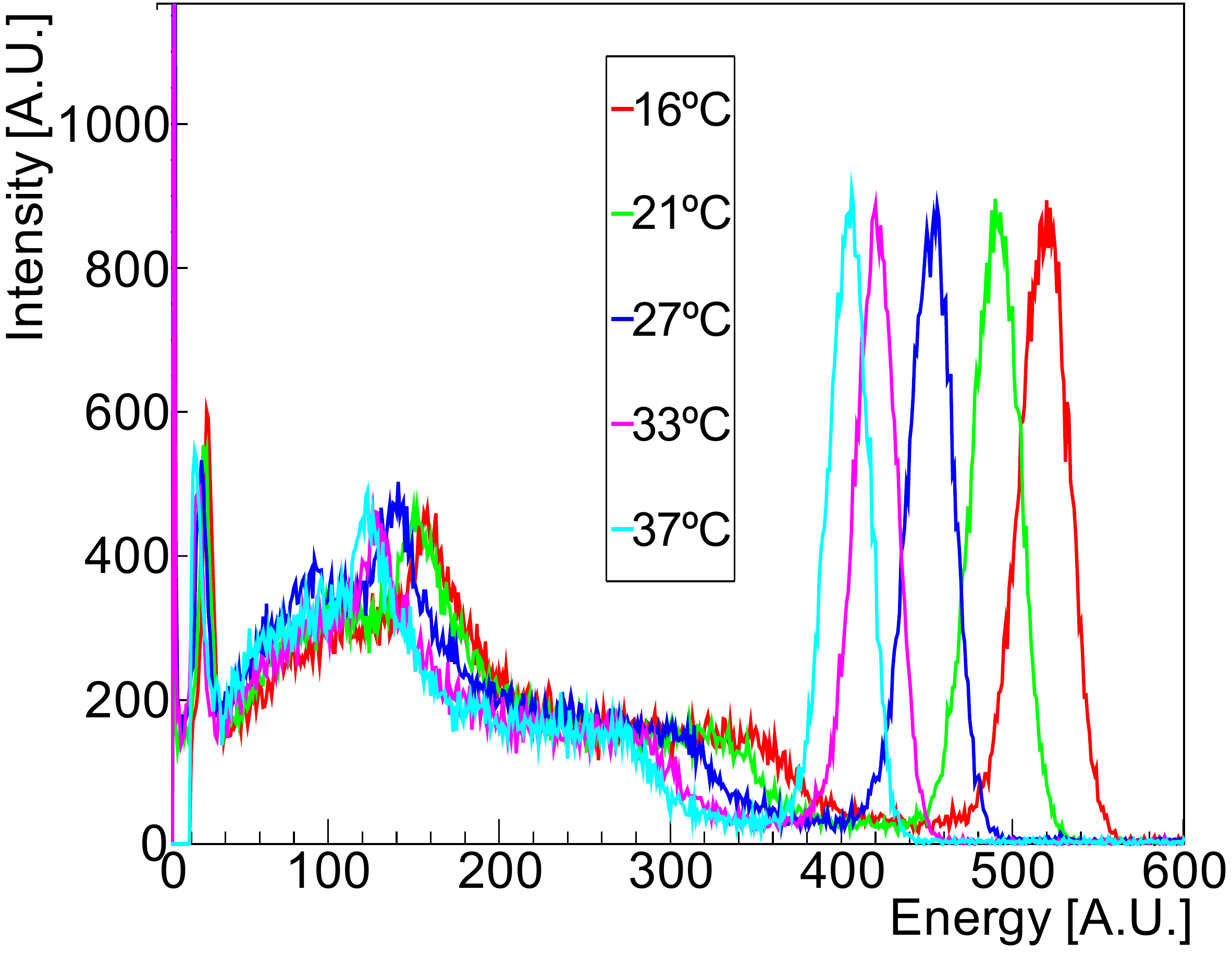}
        \caption{Uncalibrated spectra with temperature variation. As the temperature increases, the system response falls.}
        \label{fig:SpecComp}
\end{figure}

\begin{figure}[tbp]
        \centering
        \includegraphics[width=0.9\columnwidth]{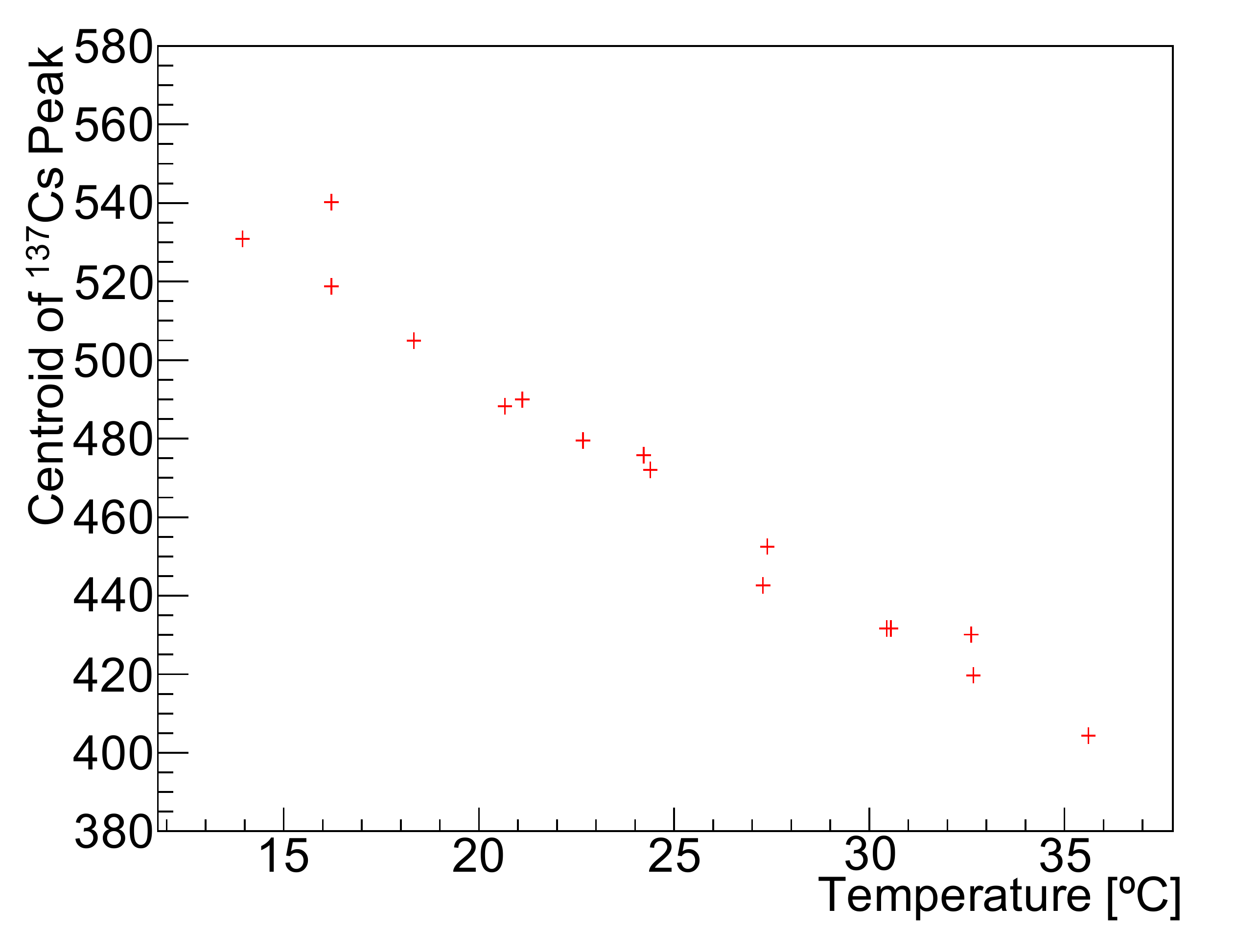}
        \caption{Uncalibrated $^{137}$Cs peaks vs. temperature. The data comes from fitting centroids to the spectra peaks, a subset of which are shown in Fig.~\ref{fig:SpecComp}.}
        \label{fig:SysTemp}
\end{figure}

We then obtained a series of datasets with the system between 14$^\circ$C and 36$^\circ$C. Fig.~\ref{fig:SpecComp} shows the spectrum acquired from a few of these datasets. A plot of the \ce{^137Cs} peak vs temperature is shown in Fig.~\ref{fig:SysTemp}. For each new temperature we allowed four hours for thermal equilibration, rather than just two, to ensure the system had fully stabilized. The system shows a clear change in the light response as the temperature increases. The decrease in the system response over the full temperature range is 24\%, of which the NaI light production decrease is 6.6\%. We attribute the remaining 17\% fall in system response to the SiPM temperature dependency, in agreement with literature values (see, e.g., Fig.2a of Ref.~\cite{Dorosz2013}).

\begin{figure}[tbp]
        \centering
        \includegraphics[width=0.9\columnwidth]{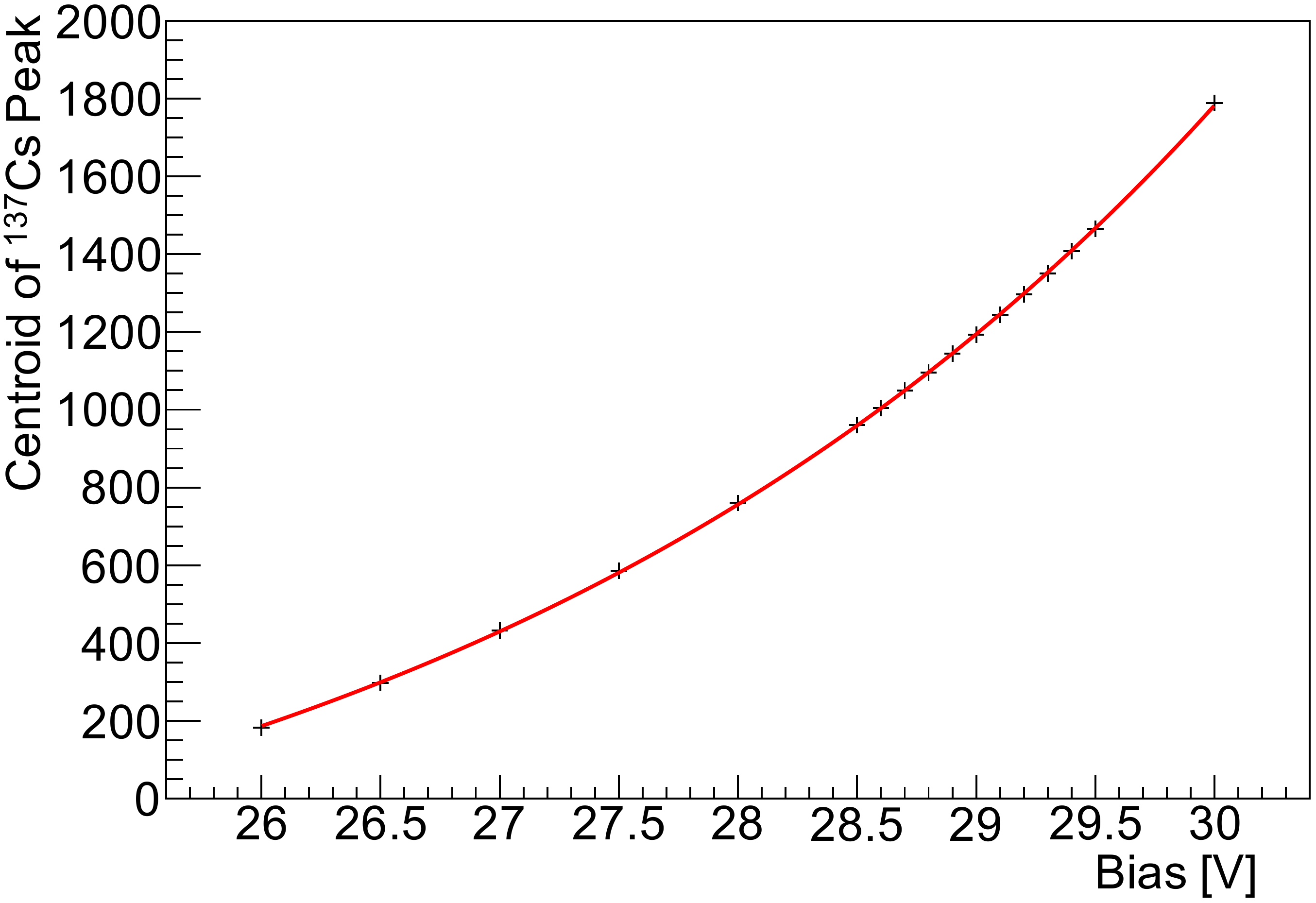}
        \caption{Uncalibrated detector response with bias variation and temperature held constant. The dynamic range of the 662~keV peak from the $^{137}$Cs source varies by 900\%. This bias-related range is 8 times larger than is required to accommodate a temperate variation of 100$^\circ$C. The empirical fit is constant value plus an exponential curve.}
        \label{fig:SiPMVoltage}
\end{figure}

We varied the bias of the SiPM array between 26~V and 30~V at room temperature to characterize its dynamic range, with the results shown in Fig.~\ref{fig:SiPMVoltage}. The system response varied by 900\% over this bias range. Given the system variation we measured of 24\% over 22$^\circ$C, this dynamic range is 8 times larger than would be required to stabilize response over a temperature change of 100$^\circ$C. We do note, however, several considerations to remain aware of in attempts to stabilize the temperature response over such a large dynamic range:

\begin{itemize}[noitemsep]
    \item The bias applied to the SiPM must have sufficient accuracy to reliably stabilize the peak centroids
    \item At lower bias, the resolution of the SiPM will worsen
    \item At lower bias, low-energy gamma ray signals, such as the 60~keV gamma rays from $^{241}$Am, may fall below the data acquisition threshold
\end{itemize}

\section{\label{sec:Usens}Uranium Enrichment Measurements}

Basic characterization of uranium samples using gamma-spectroscopy is a common in-field measurement in nuclear safeguards. In addition to the periodic background and calibration datasets, we acquired spectra from seven uranium sources with varying enrichments, four shielding configurations, and the two photodetectors. Details of the sources are given in Table~\ref{tab:sources}. The shielding configurations were:

\begin{itemize}[noitemsep]
    \item No shielding
    \item 0.635~cm steel
    \item 1.27~cm steel
    \item 1.59~cm steel
\end{itemize}

\begin{figure}[tbp]
        \centering
        \includegraphics[width=0.9\columnwidth]{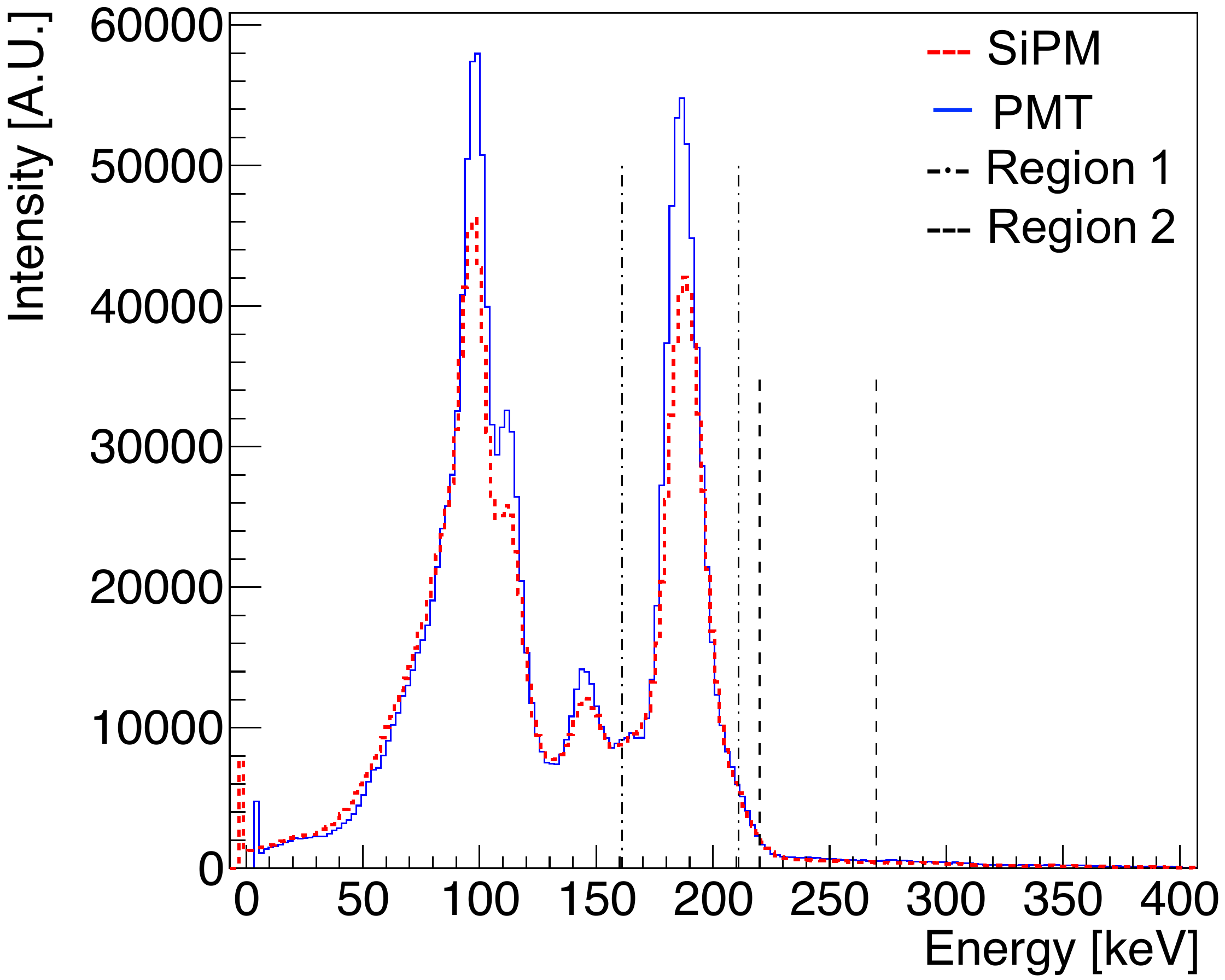}
        \caption{Comparison of the spectra of 93\% enriched uranium near the 186~keV gamma line with the PMT (solid blue) and SiPM (dashed red) mounted to the detector. Region widths used for the enrichment comparisons are shown inside the vertical black dotted and dot-dashed lines.}
        \label{fig:U93overlay}
\end{figure}

\noindent
The peak resolution at 186~keV and 1001~keV  (Fig.~\ref{fig:U93overlay}) were obtained from the unshielded 93\% enriched sample, and the resolutions are shown in Table~\ref{tab:Ures}. Note that the resolution at 1001~keV was better for the SiPM than the PMT, demonstrating the expected increase in resolution for the SiPM at high energies where the dark rate is less relevant. 

\begin{table*}[tbh!]
        \centering
        \begin{tabular}{ccc}
        \hline
        \hline
        \textbf{Source Number}   &   \textbf{Enrichment} & \textbf{Total Mass} \\
        \textbf{}   &   \textbf{$[\%]$} & \textbf{$[g]$} \\
        \hline
            1   &   93.2            &   230 \\
            2   &   52.5            &   230 \\
            3   &   20.1            &   230 \\
            4   &   4.46            &   200 \\
            5   &   2.95            &   200 \\
            6   &   0.71 (natural)  &   200 \\
            7   &   0.31 (depleted) &   200 \\
        \hline
        \hline
        \end{tabular}
        \caption{Enrichment levels and masses of the uranium sources. The masses are accurate to 0.2~g, and the enrichment levels accurate to approximately the part-per-thousand level.}
         \label{tab:sources}
\end{table*}

\begin{table*}[tbh!]
        \centering
        \begin{tabular}{rcc}
        \hline
        \hline
        \textbf{Peak Energy} & \textbf{PMT Resolution} & \textbf{SiPM Resolution}\\
        \textbf{$[keV]$} & \textbf{$[\%]$} & \textbf{$[\%]$}\\
        \hline
            186 & $8.11\pm0.04$ & $8.52\pm0.03$\\
            1001 & $6.53\pm0.84$ & $5.84\pm0.49$\\
        \hline
        \hline
        \end{tabular}
        \caption{The full-width at half-maximum energy resolution of the NaI crystal with the PMT and SiPM arrays. The resolution at 1001~keV is smaller with the SiPM-mounted detector than the PMT-mounted detector, which is the only time the SiPM performance exceeded that of the PMT.}
         \label{tab:Ures}
\end{table*}

The technique used to determine the \ce{^235U} enrichment is a linear combination of counts in the 186~keV peak and the continuum region on the high-energy side of that peak~\cite{Sprinkle1996}:

\begin{equation}
        \label{eq:energy}
        E = a\cdot S_1 + b\cdot S_2
\end{equation}

\noindent
where $S_1$ and $S_2$ are the integrated counts in Regions 1 and 2, shown in Fig.~\ref{fig:U93overlay}. Two calibration spectra are required to solve for the coefficients $a$ and $b$. The geometry of the setup for the uranium calibration sources and the unknown sources must be consistent to obtain accurate results. The samples selected for the calibration constants were sources 1 and 7. If calibration sources were chosen close to the middle of the full enrichment range (e.g., sources 2 and 3), the results were less accurate, owing to extrapolations being less reliable than interpolations. The results are shown in Table~\ref{tab:Uperc}. Each detector measures the fraction within error of each other, demonstrating comparable performance. The average accuracy of the PMT-mounted detector is ${8.5\pm6.5\%}$ and the SiPM-mounted detector is ${7.3\pm4.8\%}$.

\begin{table*}[tbh!]
    \centering
    \begin{tabular}{cccc}
        \hline
        \hline
        \textbf{Shielding} & \textbf{Enrichment} & \textbf{PMT Measured} & \textbf{SiPM Measured}\\
        \textbf{} & \textbf{$[\%]$} &\textbf{$[\%]$} & \textbf{$[\%]$}\\
        \hline
        \multirow{5}{*}{\textbf{None}} & 52.5 & $55.8\pm0.10$ & $55.6\pm0.10$ \\
            & 20.1 & $20.7\pm0.07$ & $20.3\pm0.07$ \\
            & 4.46 & $4.72\pm0.04$ & $5.21\pm0.04$ \\
            & 2.95 & $3.09\pm0.04$ & $3.45\pm0.04$ \\
            & 0.72 & $0.80\pm0.03$ & $0.76\pm0.03$\\
        \hline
        \multirow{5}{*}{\textbf{0.635~cm steel}} &52.5 & $56.1\pm0.10$ & $55.1\pm0.98$ \\
            & 20.1 & $22.0\pm0.07$ & $19.6\pm0.06$ \\
            & 4.46 & $5.15\pm0.04$ & $5.09\pm0.04$ \\
            & 2.95 & $3.40\pm0.04$ & $3.35\pm0.04$ \\
            & 0.72 & $0.74\pm0.03 $ & $0.78\pm0.03$\\
        \hline
        \multirow{5}{*}{\textbf{1.27~cm steel}} &52.5 & $55.38\pm0.10$ & $56.18\pm0.11$ \\
            & 20.1 & $20.9\pm0.07$ & $20.8\pm0.07$ \\
            & 4.46 & $5.39\pm0.05$ & $4.97\pm0.05$ \\
            & 2.95 & $3.10\pm0.04$ & $3.11\pm0.05$ \\
            & 0.72 & $0.78\pm0.04$ & $0.75\pm0.04$\\
        \hline
        \multirow{5}{*}{\textbf{1.59~cm steel}} & 52.5 & $55.33\pm0.12$ & $55.73\pm0.12$ \\
            & 20.1 & $21.0\pm0.09$ & $20.9\pm0.09$ \\
            & 4.46 & $5.69\pm0.07$ & $4.85\pm0.06$ \\
            & 2.95 & $3.05\pm0.05$ & $3.02\pm0.05$ \\
            & 0.72 & $0.76\pm0.05$ & $0.70\pm0.05$\\     
    \hline
    \end{tabular}
    \caption{Measured \ce{^235U} enrichment fraction based on the activity of the 186~keV gamma peak in multiple samples of enriched uranium. Detectors were calibrated using 93\% and depleted (0.31\%) U samples. Uncertainties are purely statistical, and any additional deviation from the known enrichment levels are attributed to systematic uncertainties. The consistency between the SiPM-mounted and PMT-mounted detectors are generally in better agreement with each other than the known enrichment values, motivating SiPMs as viable alternatives to PMTs.}
    \label{tab:Uperc}
\end{table*}


\section{\label{sec:conclusion}Summary}
  
We have discussed several disadvantages of photomultiplier tubes that possible replacement technologies could address, preferably with comparable performance. Some key traits of concern are large volume, temperature dependence, fragility, high voltage, and magnetic field dependence. 
Any replacement technology should address at least some of these concerns, while maintaining cost parity and performance with PMTs. This current work focuses on PMT replacement for medium-scale gamma ray spectrometers, with a typical dimension of 5~cm and within the context of nuclear safeguards. For the performance evaluation, our metrics are detector energy resolution, temperature compensation, and sensitivity to uranium enrichment levels. 

We performed a laboratory comparison of a PMT-instrumented and SiPM-instrumented sodium iodide detector. We calibrated the detector and measured its resolution in both cases with \ce{^241Am}, \ce{^137Cs}, and \ce{^60Co}. We found small differences in resolution between the PMT system and the SiPM system. The SiPM-mounted system exhibited sufficient dynamic range by altering the bias to compensate for the temperature-related deviations likely to be encountered in a nuclear safeguards use scenario. We further compared the resolution of the \ce{^235U} 186~keV and \ce{^238U} 1001~keV energy peaks and the results from an enrichment calculation based on the intensity of the 186~keV peak and the underlying continuum. The results were consistent with the calibration measurements at the 5-20\% level, with poorer agreement at lower enrichment levels.

SiPMs compare well to PMTs with respect to additional concerns. SiPMs are more rugged than PMTs, as they are not made of an evacuated glass bulb. The bias voltage of a SiPM is on the order of 30-100~V depending on the manufacturer and model, as compared to the 800-1500~V of a typical PMT. The SiPM is also protected against aging and accidental exposure to ambient light while fully biased, as well as being insensitive to applied magnetic fields.

\section{Acknowledgements}

Lawrence Livermore National Laboratory is operated by Lawrence Livermore National Security, LLC, for the U.S. Department of Energy, National Nuclear Security Administration under Contract DE-AC52-07NA27344. LLNL-JRNL-799528. This work was supported by the U.S. Department of Energy, National Nuclear Security Administration, Office of Nonproliferation and Arms Control, International Nuclear Safeguards Technology Development and the Human Capital Development Programs.


\end{document}